\documentclass[a4paper,twocolumn,11pt]{quantumarticle}
\pdfoutput=1
\usepackage[english]{babel}
\usepackage[utf8]{inputenc}
\usepackage[T1]{fontenc}

\usepackage{amsfonts}
\usepackage{amsmath}
\usepackage{hyperref}
\usepackage{cleveref}

\usepackage{booktabs}
\usepackage{braket}
\usepackage{mathrsfs}
\usepackage{subcaption}
\usepackage{tikz}

\crefname{equation}{}{}
\Crefname{equation}{}{}
\crefname{figure}{Figure}{Figures}
\Crefname{figure}{Figure}{Figures}
\crefname{section}{Section}{Sections}
\Crefname{section}{Section}{Sections}
\crefname{table}{Table}{Tables}
\Crefname{table}{Table}{Tables}

\newcommand{\todo}[1]{}

\begin{document}

\title{GEQIE Framework for Rapid Quantum Image Encoding}

\author{Rafał Potempa}
\orcid{0000-0002-0813-0606}
\email{rafal.potempa@polsl.pl}
\affiliation{Department of Algorithms and Software; Silesian University of Technology; Gliwice; Poland}

\author{Michał Kordasz}
\orcid{0000-0003-4193-9244}
\affiliation{MerQlab, Department of Computer Graphics, Vision and Digital Systems; Silesian University of Technology; Gliwice; Poland}

\author{Józef P. Cyran}
\affiliation{MerQlab, Department of Computer Graphics, Vision and Digital Systems; Silesian University of Technology; Gliwice; Poland}

\author{Kamil Wereszczyński}
\orcid{0000-0003-1686-472X}
\affiliation{MerQlab, Department of Computer Graphics, Vision and Digital Systems; Silesian University of Technology; Gliwice; Poland}

\author{Krzysztof Simiński}
\orcid{0000-0002-6118-606X}
\affiliation{Department of Algorithms and Software; Silesian University of Technology; Gliwice; Poland}

\maketitle

\begin{abstract}
    This work presents a Python framework named after the General Equation of Quantum Image Encoding (GEQIE). The framework creates the image-encoding state using a unitary gate, which can later be transpiled to target quantum backends. The benchmarking results, simulated with different noise levels, demonstrate the correctness of the already implemented encoding methods and the usability of the framework for more sophisticated research tasks based on quantum image encodings. Additionally, we present a showcase example of Cosmic Web dark-matter density snapshot encoding and high-accuracy retrieval (PCC = 0.995) to demonstrate the extendability of the GEQIE framework to multidimensional data and its applicability to other fields of research.
\end{abstract}
\section{Introduction}\label{sec:introduction}

Computer vision is an important part of the development of computer science. In recent years, increased interest in this field has been observed due to advances in quantum computing. The differences in data representation on a quantum computer have necessitated the development of new image-representation methods tailored to quantum computers. There are more than 30 different methods for quantum image representation (QIR) \cite{wereszczynski2023report, alwan2025comprehensive}. These include methods designed for single-channel (grayscale) images, such as FRQI \cite{le2011flexible}, NEQR \cite{zhang2013neqr},
and multichannel methods such as FRQCI \cite{li2016quantum}, IFRQI \cite{khan2019improved},
MCQI \cite{sun2013rgb}, NCQI \cite{sang2017novel}, QRCI \cite{wang2019qrci}, QUALPI \cite{zhu2021multimode} and for multidimensional images \cite{he2025quantum}.

A significant portion of quantum image representation methods belongs to the Quantum Lattice family \cite{venegas-andraca2003storing}. Each pixel is encoded using a tensor product: the pixel position and the pixel value (intensity). However, it is worth noting that there exist methods outside the Quantum Lattice family, such as OQIM \cite{xu2019orderencoded} that are not covered by the provided abstraction.

In addition to implementations of QIR methods for 2D images, attempts are made to represent 3D images. The first attempt was presented in \cite{li2014multidimensional} by introducing the NAQSS method. Later works introduce 3D Point Cloud representation \cite{jiang2017quantum}, an extension of the NEQR method for representing $n$-dimensional images \cite{semmler2023ndimensional}, and a method for 3D QIR together with a framework enabling geometric transformations \cite{he2025quantum}. In this paper, we use a multidimensional extension of the FRQI method due to its compatibility with Quantum Lattice and ease of implementation.

The growing number of QIR methods has led to some attempts to organize them by generalizing them into a universal model. In 2021 \cite{amankwah2022quantum} proposed the quantum pixel representation (QPIXL) framework. Its goal is to generalize and accelerate state preparation for existing representations by reducing the number of gates in the circuit. Another attempt is the NGQR model \cite{xing2025ngqr}, which introduces its own generalized Grayscale Quantum Coding (GQC) model along with a set of tools for its processing. The above-mentioned frameworks come with some limitations. In the case of QPIXL, the library was written in C++ and currently supports only the FRQI method. In the case of NGQR, the framework's source code is not available.
As a result of the literature analysis, an experimental gap becomes apparent: the lack of a generalized framework that would be constructed based on the Quantum Lattice scheme and written in Python, due to access to many libraries dedicated to quantum computation -- such as Qiskit \cite{javadi-abhari2024quantum} or CirQ \cite{cirqdevelopers2025cirq}.

This work presents the GEQIE library, which extends the concept introduced in \cite{potempa2024general}. The library includes a framework that has been designed with the following characteristics in mind:
\begin{itemize}
    \item open-source,
    \item usable as a Python package,
    \item equipped with an easy and ready-to-use API,
    \item based on the Quantum Lattice scheme,
    \item extendable (allowing flexible addition of methods),
    \item verifiable (enabling testing of the implemented methods).
\end{itemize}

The framework was implemented in Python using the Qiskit library and equipped with an API interface that allows it to be invoked as a module in any project. The encoding is based on the Quantum Lattice scheme, which underlies many QIR methods. An additional feature of the framework is the ability to add new methods and test their correctness. The code is available as an open-source repository in GitHub under \href{https://github.com/merQlab/geqie}{github.com/merQlab/geqie}.

This library was used in a web-based playground that allows running simulations and testing new and modified image encoding methods. The application has been made public, and its link can be found in the GitHub repository.

\cref{sec:materials_and_methods} presents the mathematical description of the GEQIE model and the benchmarking metrics, later used for experimental purposes. \cref{sec:results} shows and describes the results of the experiments, both for the grayscale and RGB images and for the simulation of the Cosmic Web. \cref{sec:discussion} focuses on the discussion of the results provided, while \cref{sec:conclusions} highlights key findings of the article. 

\section{Materials and methods}\label{sec:materials_and_methods}

\subsection{General Quantum Image Representation Model}\label{sec:general_quantum_representation}

We consider a digital image as a 2D array of pixels with $C$ channels (e.g., $C=1$ for grayscale, $C=3$ for RGB). 
Let $H$ and $W$ denote the height and width of the image, respectively. Each pixel is indexed by coordinates $(u,v)$ with 
$u\in\{0,\dots,H-1\}$ and $v\in\{0,\dots,W-1\}$, and stores a channel vector
$p_{uv}\in[0,1]^C$ (values assumed normalized to $[0,1]$ per channel).

To build a unified description of quantum image representations, we separate two conceptual roles:

\begin{itemize}
    \item \textbf{position encoding}: a mapping that associates the coordinates of the pixel $(u,v)$ with a quantum state that identifies this pixel within the global superposition,
    \item \textbf{value encoding}: a mapping that converts the pixel intensity/color $p_{uv}$ (possibly multi-channel) into a quantum state or amplitude pattern.
\end{itemize}

\paragraph{Position map.}
We introduce a map
\begin{equation}\label{eq:xi_defn}
    \xi:\mathbb{Z}\times \mathbb{Z} \to \mathscr{F},
\end{equation}
where $\mathscr{F}$ is a chosen set of quantum states used to represent pixel positions. The only essential requirement is
\emph{decodability}: from the state $\xi(u,v)$ one must be able to recover the coordinates $(u,v)$ (i.e., $\xi$ is injective on valid pixel coordinates).
A typical case is the \emph{qubit lattice} setting, where $\mathscr{F}$ is the computational basis of an index register and $\xi(u,v)$ is simply a basis state encoding row/column indices.

\paragraph{Value map.}
Let $\mathscr{D}$ denote the Hilbert space used to represent pixel values (intensity or color).
Its dimension is typically $2^D$, where $D$ is the number of qubits allocated to value encoding.
We write the value map as
\begin{equation}\label{eq:delta_u_v_c}
    \delta : \mathbb{Z}\times\mathbb{Z}\times[0,1]^C \to \mathscr{D}.
\end{equation}
In practice, $\delta$ is defined only for valid pixel coordinates; when an implementation uses power-of-two registers,
we conceptually embed the $H\times W$ image into a larger $N_H\times N_W$ grid and set the out-of-range contributions to zero, which is the standard way to handle non-power-of-two image sizes using qubit registers.

An important limiting case is $D=0$, where the value is represented directly by a complex amplitude, i.e., $\mathscr{D}=\mathbb{C}$. This recovers amplitude-encoding-like models.

\paragraph{General Equation of Quantum Image Encoding (GEQIE).}
Using these two components, the general quantum state representing an image is written as
\begin{widetext}
\begin{equation}\label{eq:general_quantum_image_model}
    \ket{I_G(\delta,\xi)}_k
    =
    \frac{1}{\sqrt{HW}}
    \sum_{u=0}^{N_H-1}\sum_{v=0}^{N_W-1}
    \sum_{l=0}^{L-1}
    \delta_{k,l}(u,v,p_{uv})\otimes \xi_{k,l}(u,v),
\end{equation}
\end{widetext}
where $k\in\{1,\dots,K\}$ indices separate (non-interfering) components used by some schemes (e.g., multi-register or multi-cluster encodings) -- these components may still belong to one physical system, e.g., via cluster/graph connectivity, but are treated as distinct state blocks in the representation, $l\in\{0,\dots,L-1\}$ allows an additional summation layer present in some models (if not needed, one can set $L=1$ and drop $l$), $N_H$ and $N_W$ are powers of two determined by the available index registers, with $N_H\ge H$ and $N_W\ge W$, out-of-range coordinates ($u\ge H$ or $v\ge W$) contribute zero by construction of $\delta$, so only $HW$ terms remain effective.

\paragraph{Normalization.}
Although the formal sums run over $N_HN_W$ indices, the value map cancels all out-of-image positions. Therefore, the state contains
$HW$ nonzero pixel contributions, and the natural normalization factor is $1/\sqrt{HW}$.

\paragraph{Quantum image encoding model.}
We call
$Q=(K,L,D,\{\delta_{k,l}\},\{\xi_{k,l}\})$ a \emph{quantum image encoding model} if it specifies:
(i) the number of components $K\ge 1$,
(ii) the optional extra summation $L\ge 1$ (or equivalently $L=1$ when unused),
(iii) the number of value qubits $D\ge 0$,
and (iv) concrete maps $\delta_{k,l}$ and $\xi_{k,l}$ that instantiate \cref{eq:general_quantum_image_model}.

\begin{figure}[htbp]
    \centering
    \includegraphics[width=\linewidth]{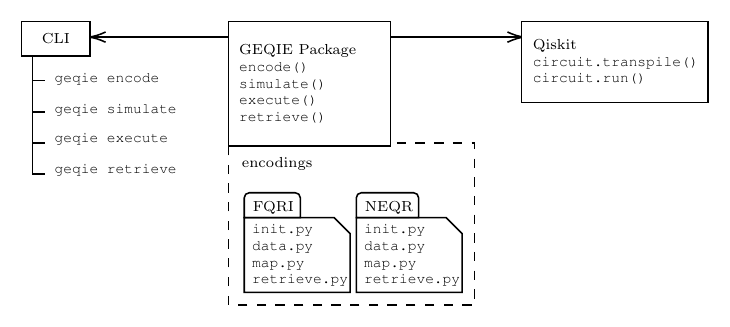}
    \caption{The schematic diagram of the GEQIE framework. From the left are visible the Command Line Interface used, e.g., by the web application; the GEQIE package with encoding methods containing init, data, and map associated with \cref{eq:general_quantum_image_model} plus the retrieve function; and the interaction with the Qiskit framework that covers automatic transpilation of the circuits.}
    \label{fig:geqie_schematic}
\end{figure}

The application of the presented quantum image encoding model in the GEQIE library is presented in the scheme in \cref{fig:geqie_schematic}.

\subsection{Benchmarking metrics}

\subsubsection{Pearson's Correlation Coefficient}

The metric used to measure image similarity is Pearson's Correlation Coefficient (PCC). The use of this metric for this task is well established in the field \cite{starovoitov2020comparative}. The formula used to calculate it is
\begin{equation}\label{eq:pcc}
    \text{PCC} = \frac{ \sum(x-m_x) (y-m_y) }{ \sqrt{ \sum(x-m_x)^2 \sum(y-m_y)^2} },
\end{equation}
where $x$ and $y$ are the data vectors (flattened image pixel values), and $m_x$ and $m_y$ are mean values of the respective vectors \cite{virtanen2020scipy}.

\subsubsection{Peak Signal-to-Noise Ratio}

The Peak Signal-to-Noise Ratio (PSNR) is widely used to assess the quality of a signal relative to noise. In this paper, we treat the signal as the original image pixel values, and the noise comes from the intrinsic properties of the noisy simulation, performed as described in \cref{scn:noise_simulation}. The formula used for PSNR calculation is
\begin{equation}\label{eq:psnr}
    \text{PSNR} = 20 \log_{10}\left( \frac{255}{\sqrt{\text{MSE}_{xy}}} \right),
\end{equation}
where the value $255$ is taken as the maximum value of an 8-bit resolution of a color channel, and $\text{MSE}_{xy}$ is the Mean Squared Error of the flattened vector pixel values of retrieved and original images $x$ and $y$. For perfect retrieval ($\text{MSE}_{xy} = 0$), we indicate the PSNR value as \emph{inf} for simplicity.

\subsection{Noise simulation}\label{scn:noise_simulation}

We model noise as a completely positive trace-preserving (CPTP) channel acting after the ideal unitary evolution. While our framework constructs the algorithm as a single global unitary, noise can be introduced as a tensor product of single-qubit depolarizing channels. This approach follows the standard abstraction used both in fault-tolerant and NISQ simulations, where ideal logical evolution and physical noise processes are treated separately \cite{isakov2021simulations}.

The depolarizing quantum error used in benchmarking is defined as
\begin{equation}\label{eq:noise_aer}
    E(\rho) = (1 - \lambda) \rho + \lambda \text{Tr}[\rho] \frac{I}{2^n},
\end{equation}
where $\rho$ is the density matrix, $\lambda$ is error probability, $n$ is the number of qubits \cite{javadi-abhari2024quantum}. In this paper, we use $\lambda$ values from $\{0, 0.01, 0.1, 0.2, 0.5, 0.9, 1.0\}$, which should cover all the interesting cases like ideal, low, and heavy noise, and also the destructive regime.

Full circuit noise simulation is a resource-intensive task, especially for methods that require more qubits. For larger circuits, machines available on the Google Cloud Platform were used, with the upper limit of 864 GB of RAM (n2-highmem-128). Samples that exceeded 12 qubits (NEQR $8\times8$, QUALPI $8\times8$, NCQI $4\times4$, and $8\times8$) were omitted from the benchmarking due to the high memory consumption of the circuits.

\subsection{Cosmic Web simulations}

Cosmic Web simulations are a broad branch of cosmology that focuses on validating cosmological models such as $\Lambda$CDM \cite{aghanim2021planck}. Over the years, simulations have run on different scales, ranging from 1000 objects \cite{press1974formation} to complex, modern, long-running simulations such as Millennium-II \cite{springel2005simulations} and TNG-Cluster \cite{nelson2024introducing}, with more than $5\cdot10^{12}$ dark matter particles, and have also considered magnetic and hydrodynamic effects. As cosmological simulations grow in size, the demand for computing power increases. The computational complexity of some approximate simulation algorithms is $\mathcal{O}(\log(n))$ \cite{tassev2013solving, howlett2015lpicola}. The approaches mentioned use simplifications that come at a slight loss in simulation accuracy; therefore, studying quantum alternatives is essential to identify problem instances or regimes where quantum representations may offer advantages.

To overcome this issue, other paradigms than classical n-body simulations are used, such as neural networks \cite{bonici2025effort} and initial trials of the quantum computing approach \cite{mocz2021cosmological}. The latter is the most interesting to us; thus, in this paper, we present a toy example of encoding Cosmic Web simulation results as a quantum image to verify and discuss its potential to speed up cosmological simulations, while also demonstrating the framework's extendibility to higher-dimensional data. The results of these experiments can be found in \cref{scn:cosmic_web_image_encoding}.

\subsection{Web playground}

\begin{figure}
    \centering
    \includegraphics[width=0.6\linewidth]{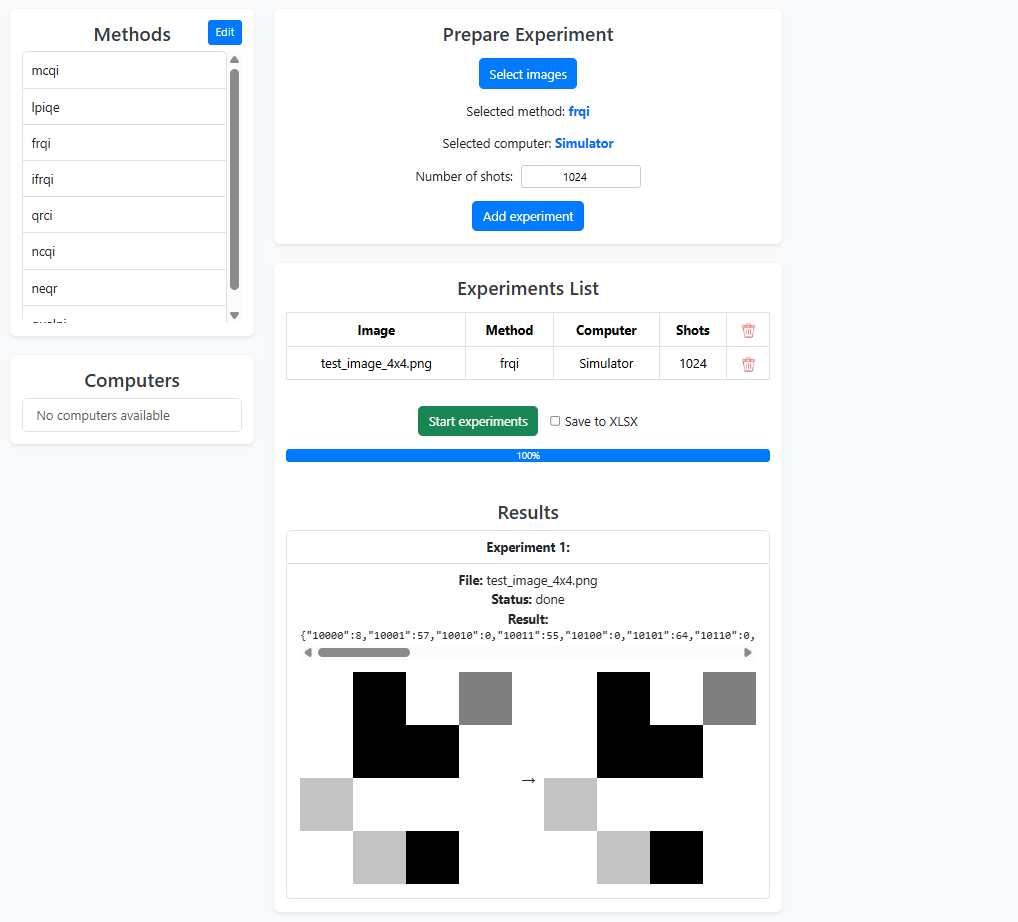}
    \caption{GEQIE GUI with available list of quantum encoding methods, possible list of devices to execute the circuit on, and experimental setup in the middle.}
    \label{fig:geqie_gui}
\end{figure}

The web application uses the GEQIE framework to allow limited experimentation with existing methods and their modification for learning purposes. The example view of the web application is shown in \cref{fig:geqie_gui}.

\section{Results} \label{sec:results}

 In this section, we present the results of benchmarking different 2D image encodings on randomly generated grayscale and RGB images. The section also contains a toy example of experimentation on Cosmic Web simulation results, encoding them into a multidimensional version of the FRQI method, together with a discussion of its potential usefulness for the Cosmic Web simulations field.

The generated data and benchmarking code are available in the GEQIE repository on GitHub.

\subsection{2D image encoding}

Currently, the framework supports eight Quantum Image Encoding methods: four grayscale and four RGB. Each method was tested against three image sizes, i.e., $2\times2$, $4\times4$, and $8\times8$ pixels, eight images each. The ideal simulation is complemented with simplified noisy simulations, as described in \Cref{scn:noise_simulation}.

During benchmarking, each retrieved image was tested against the original image with Pearson's Correlation test \cref{eq:pcc} to verify its likelihood with the original, as well as the Peak Signal-to-Noise Ratio (PSNR) \cref{eq:psnr} to check the method's encoding robustness against the noisy scenario. 

The raw values are presented in \cref{tab:pcc_table} and \cref{tab:psnr_table} for PCC and PSNR, respectively. Especially for smaller images and more accurate methods, e.g., NEQR, QUALPI, and NCQI, the obtained results were perfect with respect to the original image, even for noisy scenarios. These methods use a large number of qubits to represent pixel color, which yields high accuracy in the final representation; however, this results in a resource-intensive simulation. In some scenarios, we were unable to perform noisy or ideal simulations for larger image sizes (NEQR, QUALPI, NCQI), so those scenarios are missing from the results. The results with PCC scores of 1.0 are represented in \cref{tab:psnr_table} as \emph{inf} for simplicity, since the image error and MSE are zero with a perfect match.

\begin{table}[htbp]
    \centering
    \includegraphics[width=0.48\textwidth]{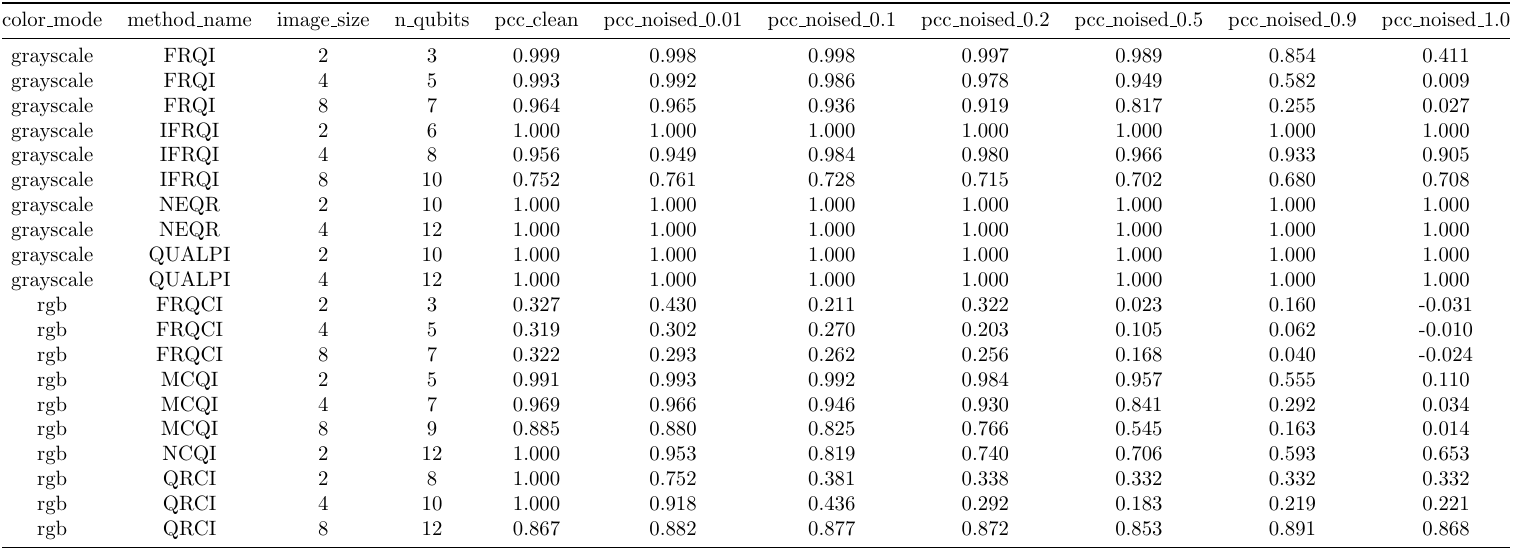}
    \caption{Average values of Pearson's Correlation Coefficient of retrieved vs original image.}
    \label{tab:pcc_table}
\end{table}

\begin{table}[htbp]
    \centering
    \includegraphics[width=0.48\textwidth]{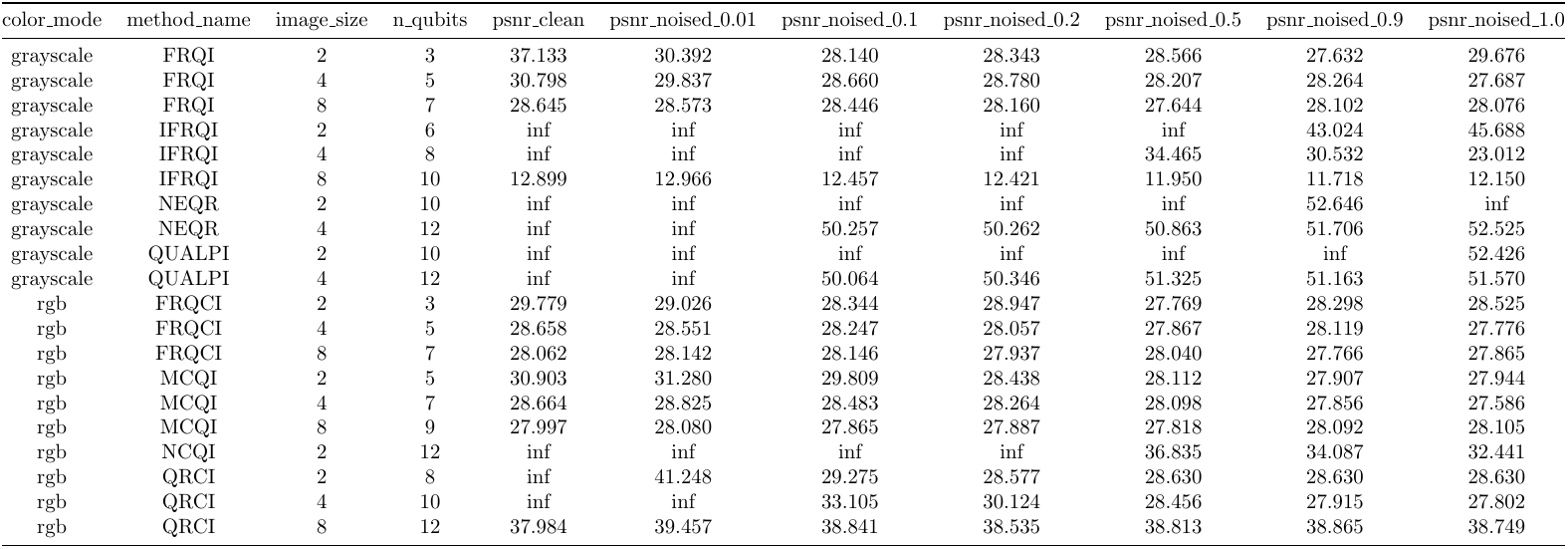}
    \caption{Average value of Peak Signal-to-Noise Ratio (PSNR) of retrieved vs original image.}
    \label{tab:psnr_table}
\end{table}

\begin{figure}[htbp]
    \centering
    \begin{subfigure}{0.23\textwidth}
        \includegraphics[width=\textwidth]{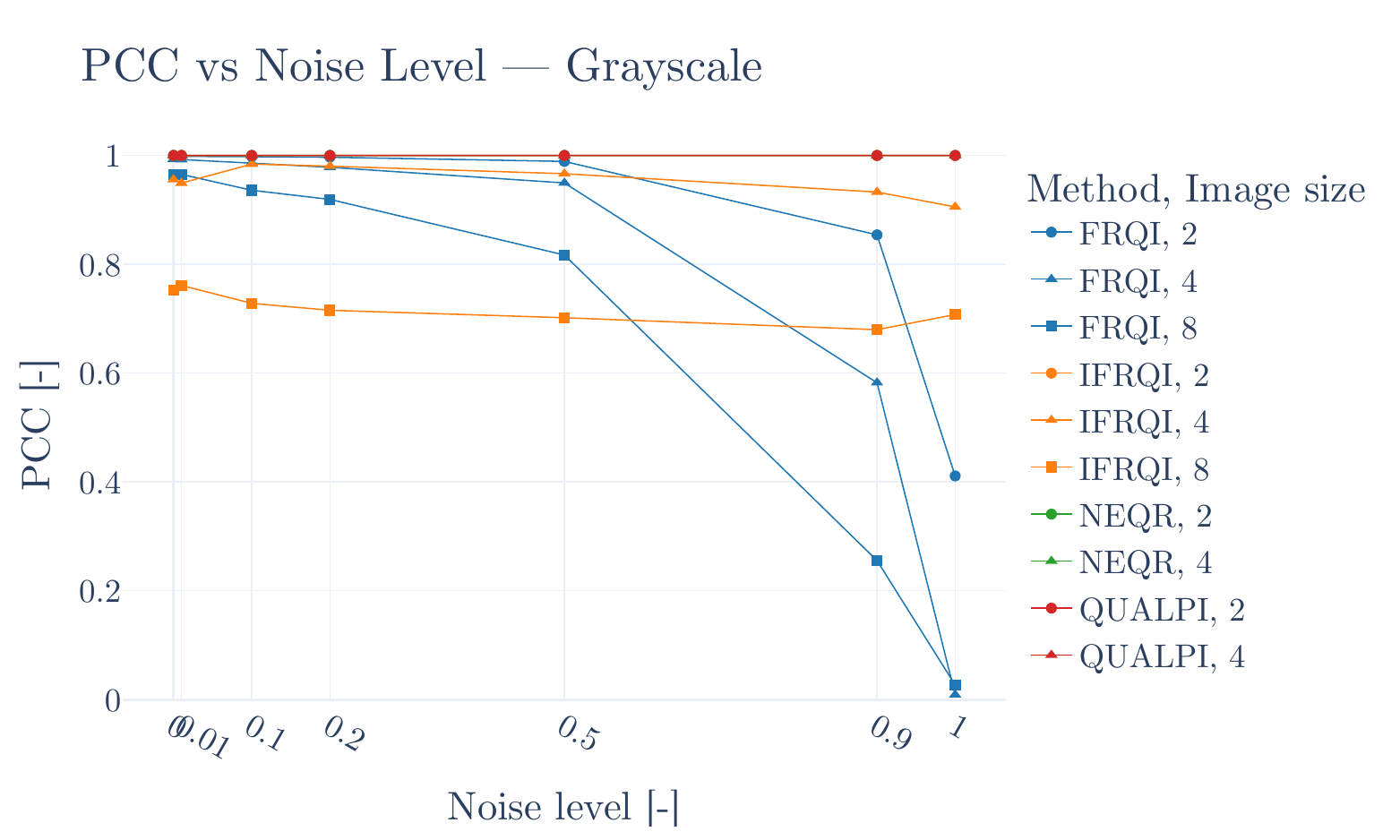}
        \caption{grayscale}
        \label{fig:pcc_vs_noise_grayscale}
    \end{subfigure}
    \begin{subfigure}{0.23\textwidth}
        \includegraphics[width=\textwidth]{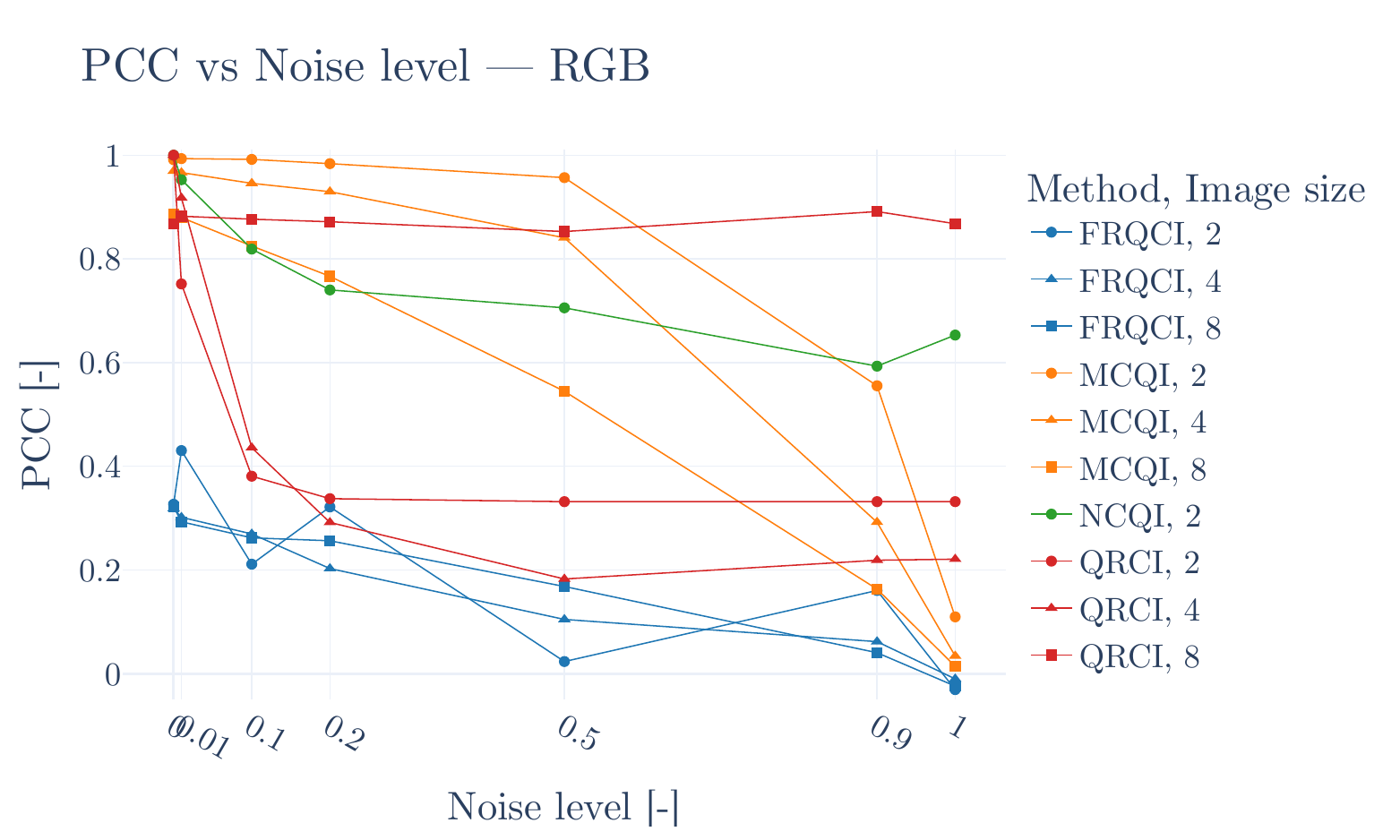}
        \caption{RGB}
        \label{fig:pcc_vs_noise_rgb}
    \end{subfigure}

    \caption{Average value of Pearson's Correlation Coefficient of retrieved vs original image, versus noise level for (a) grayscale and (b) RGB methods.}
    \label{fig:pcc_vs_noise}
\end{figure}

\begin{figure}[htbp]
    \centering
    \begin{subfigure}{0.23\textwidth}
        \includegraphics[width=\textwidth]{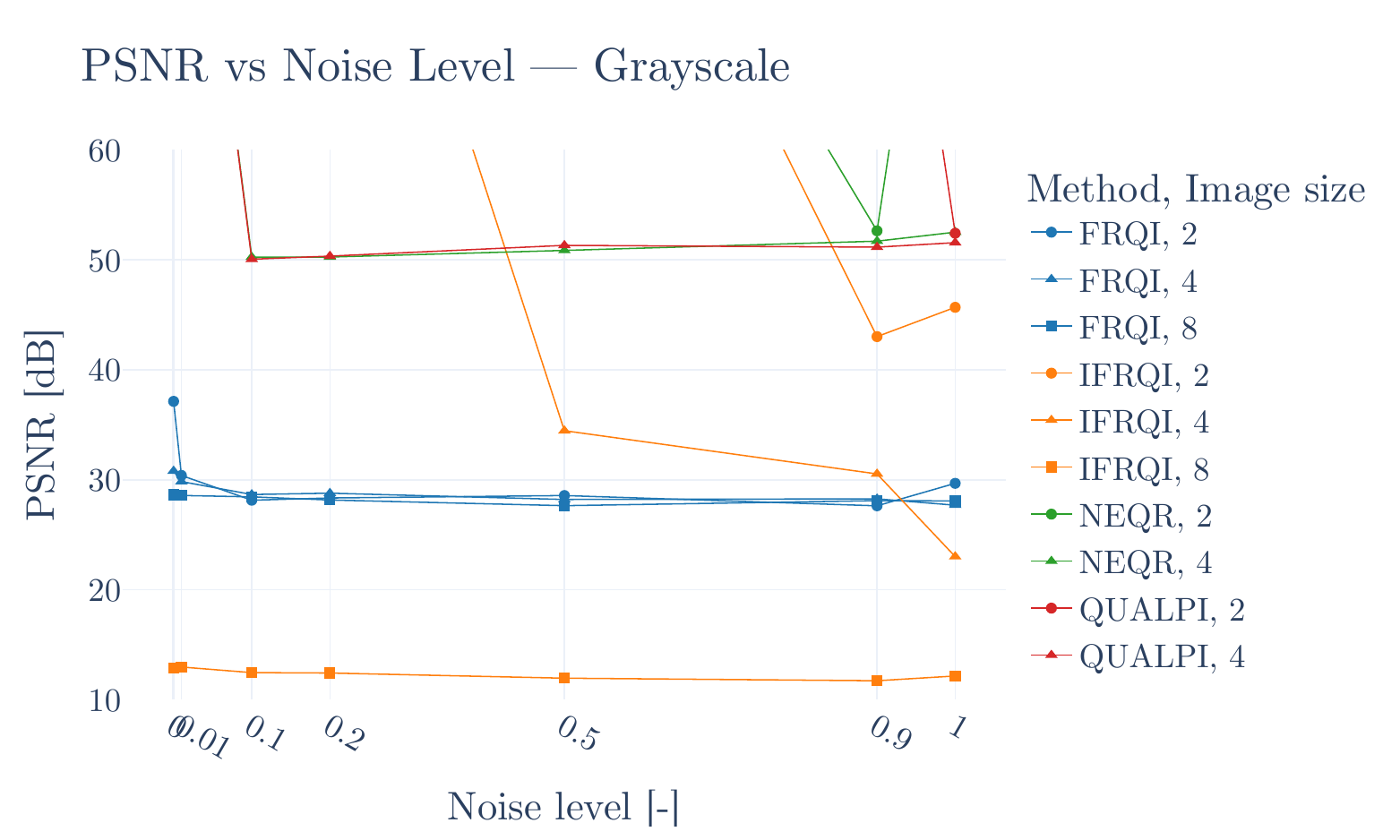}
        \caption{grayscale}
        \label{fig:psnr_vs_noise_grayscale}
    \end{subfigure}
    \begin{subfigure}{0.23\textwidth}
        \includegraphics[width=\textwidth]{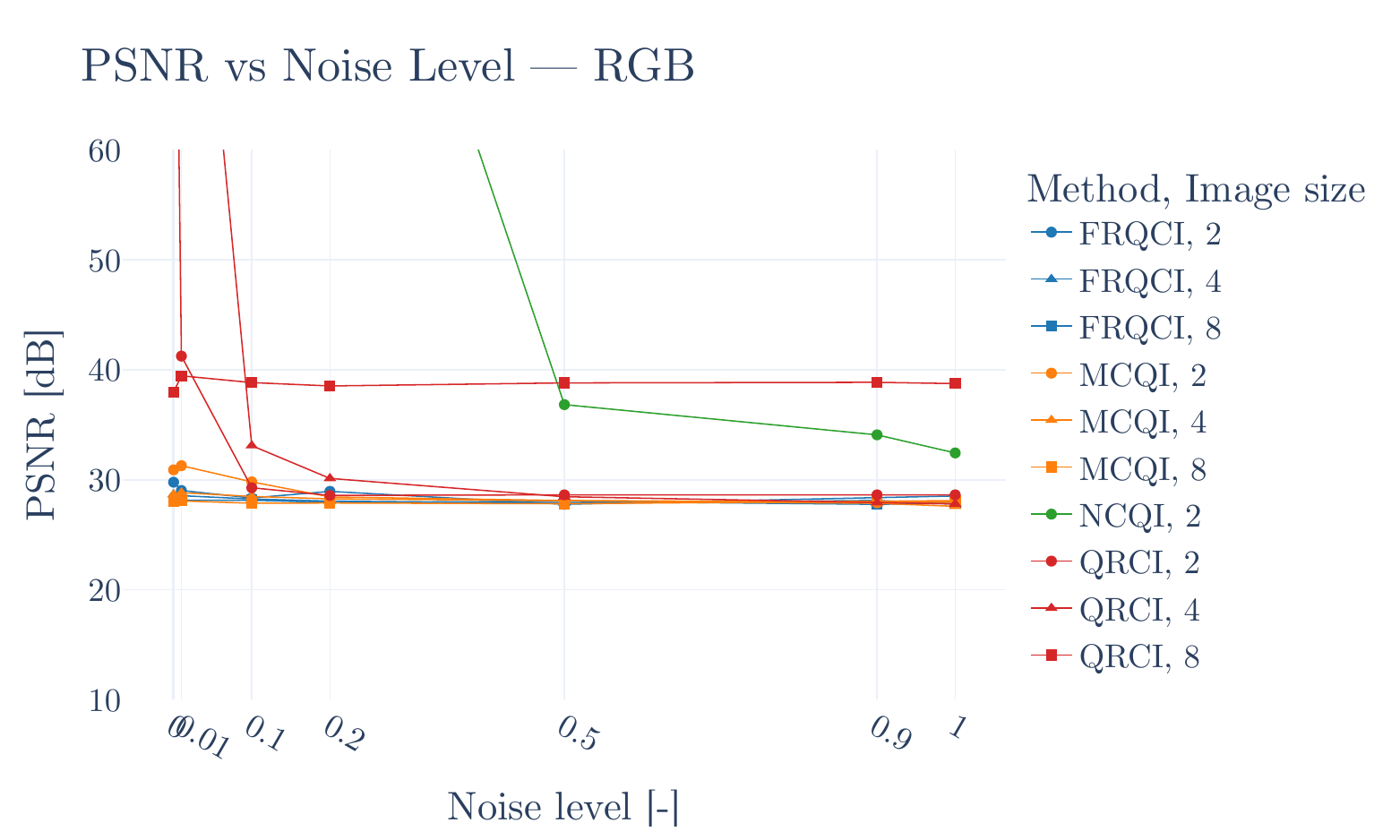}
        \caption{RGB}
        \label{fig:psnr_vs_noise_rgb}
    \end{subfigure}

    \caption{Average value of Peak Signal-to-Noise Ratio (PSNR) of retrieved vs original image, versus noise level for (a) grayscale and (b) RGB methods. For perfect retrievals, the MSE used in the PSNR calculation was zero, and the assumed PSNR was infinite; the values are deliberately capped at 60 dB to indicate this.}
    \label{fig:psnr_vs_noise}
\end{figure}

The plots in \cref{fig:pcc_vs_noise} and \cref{fig:psnr_vs_noise} represent the PCC and PSNR versus the noise value $\lambda$ as defined in \cref{eq:noise_aer}. From \cref{fig:pcc_vs_noise_grayscale}, it is clear that methods that use more qubits for color encoding, like IFRQI, NEQR, and QUALPI, achieve better accuracy without noise and also preserve encoding quality in the noisy region, while the basic method, like FRQI, loses the data completely. 

In the RGB methods shown in \cref{fig:pcc_vs_noise_rgb}, it is evident that FRQCI fails to grasp images correctly, even in an ideal simulation. The error comes from the color encoding mechanism, which results in the blue channel being completely obscured by the sampling error. To fix this, a high number of shots is required. MCQI shows good retrieval quality in ideal simulations but degrades in noisy scenarios. QRCI with $8\times8$ shows the best noise resilience, while NCQI $2\times2$ also gives promising results, which should, however, be extended to higher image resolutions.

\subsection{IMB Quantum Platform execution}

\begin{figure}[htbp]
    \centering
    \includegraphics[width=0.4\textwidth]{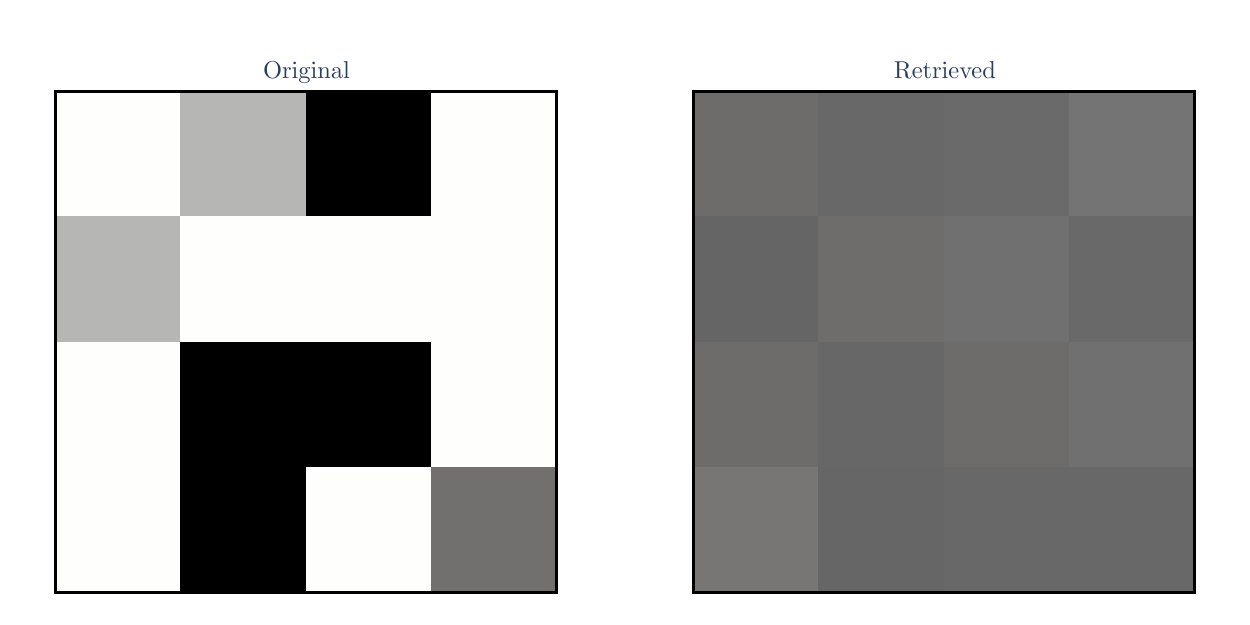}
    \caption{Comparison of the original and retrieved images encoded with the FRQI method using GEQIE and executed on the IBM Quantum Platform. The relative metrics of the images are low due to high noise (PCC=0.474, PSNR=30.98 dB).}
    \label{fig:execute_frqi_retrieved_image}
\end{figure}

The \texttt{geqie execute} functionality allows deploying the encoded circuit to the IBM Quantum Platform using an API token or credentials stored in the system. To verify this functionality, we executed one $4\times4$ grayscale image encoded using FRQI and NEQR methods on IBM Heron r2, with 4096 shots. The results shown in \cref{fig:execute_frqi_retrieved_image} indicate that the real noisy environment scrambles the results, suggesting that error-correction techniques should be used on NISQ devices. Nevertheless, recent results on error correction might suggest that we are leaving the NISQ era towards Fault-Tolerant computations \cite{acharya2025quantum} using logical qubits. 

\subsection{Cosmic Web image encoding}\label{scn:cosmic_web_image_encoding}

\begin{figure}[htbp]
    \centering
    \includegraphics[width=\linewidth]{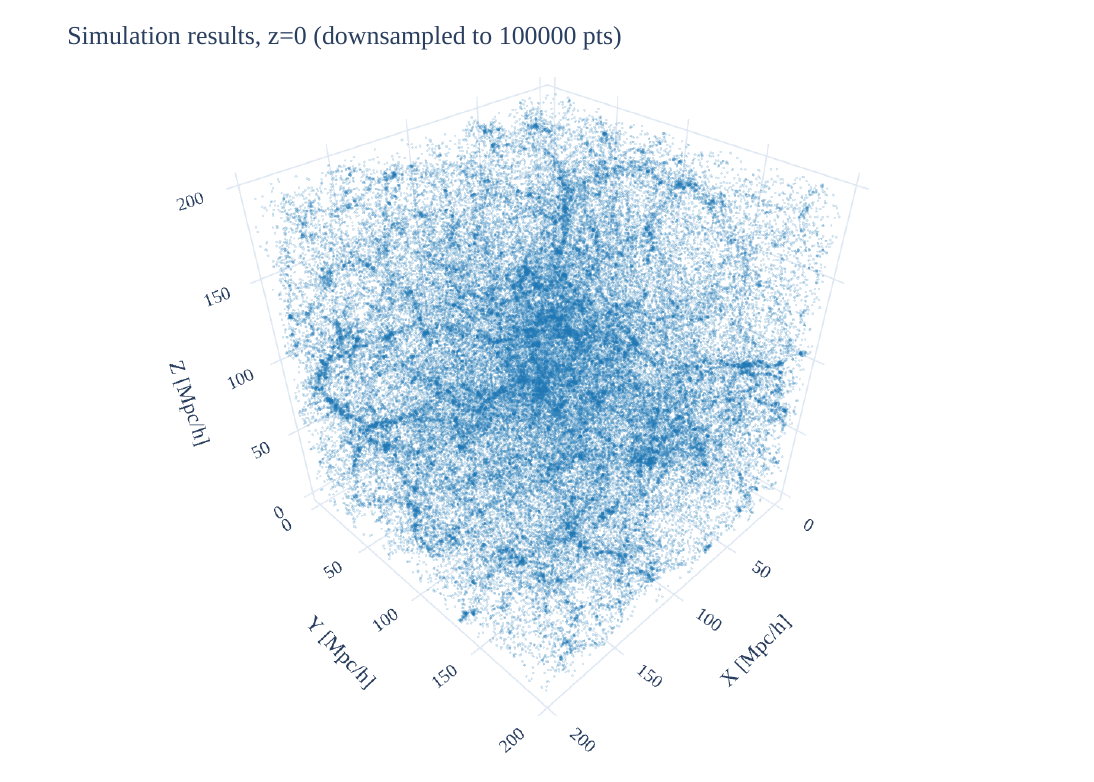}
    \caption{L-PICOLA simulation results down-sampled to $10^5$ points. The snapshot contains ~$2\cdot10^6$ particles and has a size of 200 Mpc/h.}
    \label{fig:snapshot_3d_scatter}
\end{figure}

In this work, we use L-PICOLA \cite{howlett2015lpicola} simulation with default parameters, to work on quantum encoding of the 3D point cloud snapshot, shown in \cref{fig:snapshot_3d_scatter}. The input files and data used for this analysis are available in the GitHub repository.

The idea here was to encode the mass distribution of the n-body results into a voxel grid of a specific resolution. For this, a normalizing function has been selected that maps the mass distribution to the range $[0, 1]$ as uniformly as possible, while remaining a simple and invertible mathematical function. 

The explored normalization function candidates are:
\begin{itemize}
    \item Weibull-like normalization with mean $\bar{x}$
        \begin{equation}\label{eq:normalizaion_e_mean}
            1 - e^{-x/\bar{x}},
        \end{equation}
    \item Weibull-like normalization with median $\tilde{x}$
        \begin{equation}\label{eq:normalizaion_e_median}
            1 - e^{-x/\tilde{x}},
        \end{equation}
    \item Base-10 normalization with mean $\bar{x}$
        \begin{equation}\label{eq:normalizaion_10_mean}
            1 - 10^{-x/\bar{x}},
        \end{equation}
    \item Base-10 normalization with median $\tilde{x}$
        \begin{equation}\label{eq:normalizaion_10_median}
            1 - 10^{-x/\tilde{x}}.
        \end{equation}
\end{itemize}

\begin{figure}[htbp]
    \centering
    \includegraphics[width=\linewidth]{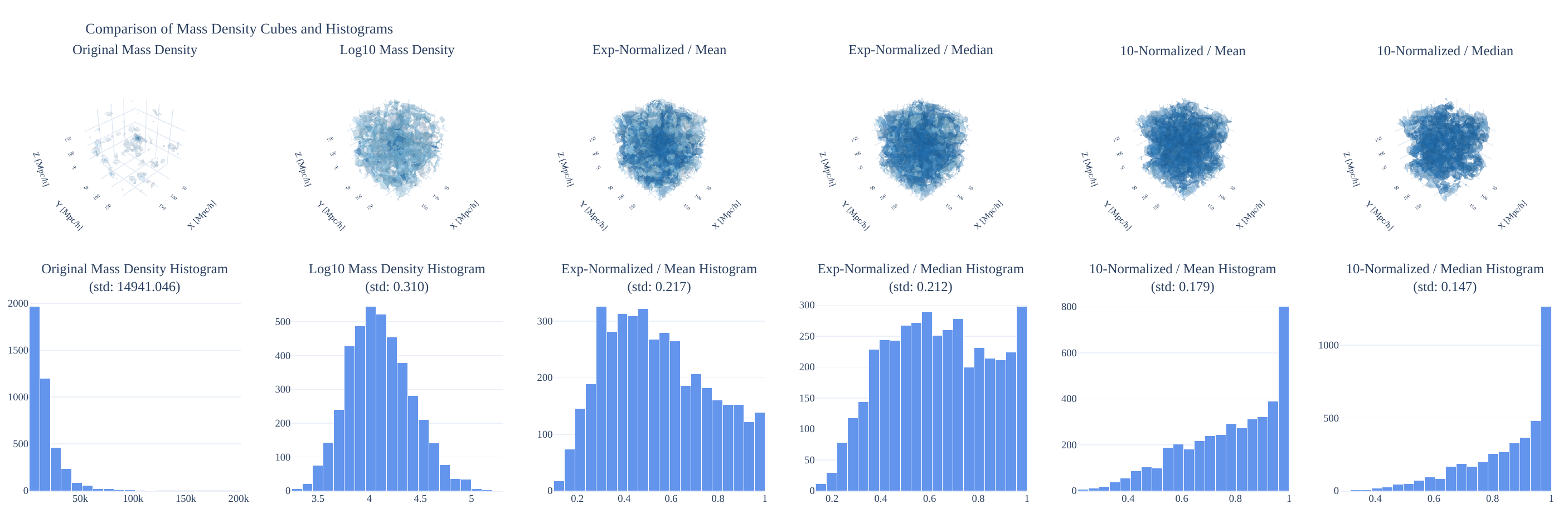}
    \caption{Low resolution ($16^{\times3}$) normalization summary with (up) mesh isosurface visualization and (down) histograms of data distribution, with the first two left sets resembling original data and $\log_{10}$ normalized.}
    \label{fig:summary_low_resolution}
\end{figure}

The first trial used a low-resolution voxel grid of size $16^{\times3}$, with the summary shown in \cref{fig:summary_low_resolution}, where we additionally used the standard deviation as a metric to quantify the data spread. This indicated the best normalization method being either \cref{eq:normalizaion_e_mean} or \cref{eq:normalizaion_e_median}, with \cref{eq:normalizaion_e_mean} having slightly bigger spread ($\sigma=0.217$ vs $\sigma=0.212$). Another fact that might support the use of the mean is the computational complexity of calculating the median. The 10-based normalization should be disregarded due to the high concentration of large densities near 1.0, which could lead to greater information loss when subjected to image retrieval sampling error.

\begin{figure}[htbp]
    \centering
    \includegraphics[width=\linewidth]{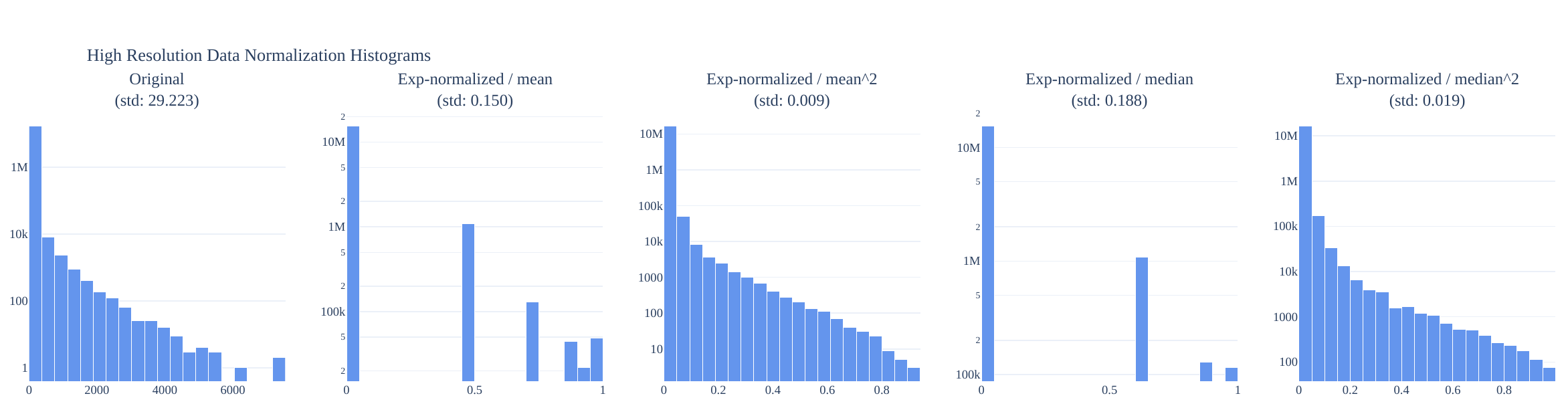}
    \caption{High resolution ($256^{\times3}$) normalization histograms summary.}
    \label{fig:summary_high_resolution}
\end{figure}

The second experiment -- with higher resolution grid division ($256^{\times3}$) -- shows important issues: vast majority of voxels is empty, thus skewing mean and median towards zero, and also showing quantization of density, i.e., smaller voxel contain only zero, one, two, three, etc. objects of the same mass, leaving gaps in the resulting distribution. One method to overcome this is a heavier scaling factor, for which the nonzero voxel means, medians, and their squares have been tested. The results shown on \cref{fig:summary_high_resolution} make it clear that $\sigma$ no longer meets its role due to the skew towards zero, however distributions are visually more uniform for squared values of mean $\bar{x}_{\neq0}$ and median $\tilde{x}_{\neq0}$ computed over nonzero elements only. Hence, we suggest using either of those methods for the higher resolution snapshot normalization:
\begin{itemize}
    \item Weibull-like normalization with nonzero elements mean $\bar{x}_{\neq 0}^2$
        \begin{equation}\label{eq:normalizaion_e_mean_squared}
            1 - e^{-x/\bar{x}_{\neq0}^2},
        \end{equation}
    \item Weibull-like normalization with nonzero elements median $\tilde{x}_{\neq 0}^2$
        \begin{equation}\label{eq:normalizaion_e_median_squared}
            1 - e^{-x/\tilde{x}_{\neq0}^2},
        \end{equation}
\end{itemize}

To verify the usefulness of this method, we implemented the multidimensional version of the FRQI method (MFRQI) and tested it on a low-resolution sample. The sample was normalized using \cref{eq:normalizaion_e_median}, encoded using the MFRQI method in GEQIE, simulated with a high amount of shots ($2^{20}$) for the best output quality, and later de-normalized using a function inverse to \cref{eq:normalizaion_e_median}.

\begin{figure}[htbp]
    \centering
    \begin{subfigure}{0.23\textwidth}
        \includegraphics[width=\textwidth]{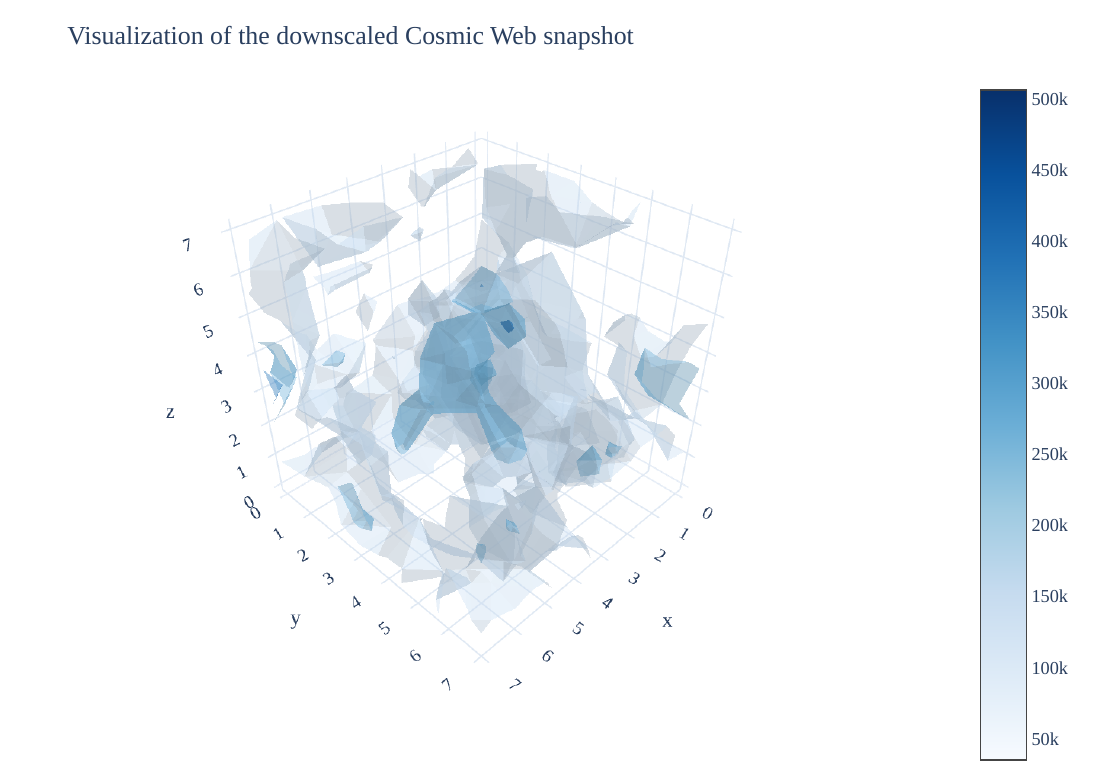}
        \caption{original}
        \label{fig:mfrqi_original_snapshot}
    \end{subfigure}
    \begin{subfigure}{0.23\textwidth}
        \includegraphics[width=\textwidth]{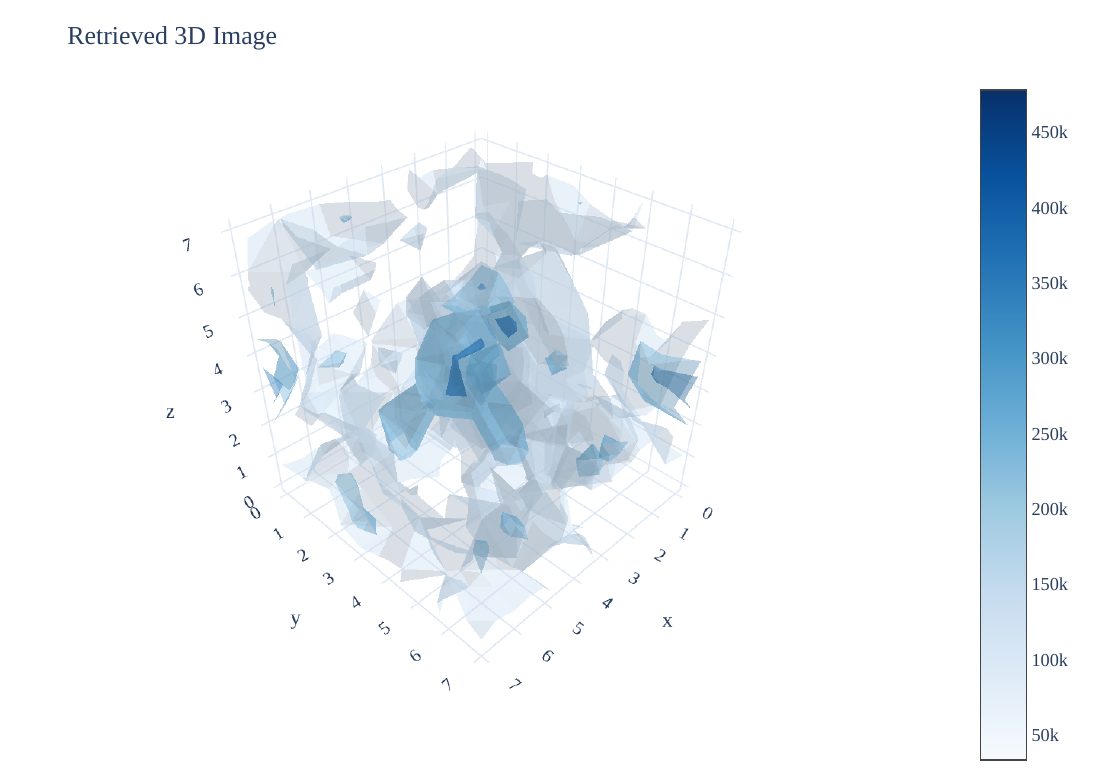}
        \caption{retrieved}
        \label{fig:mfrqi_retrieved_snapshot}
    \end{subfigure}

    \caption{Comparison of (a) original vs (b) retrieved snapshot generated using GEQIE framework (PCC=0.995).}
    \label{fig:mfrqi_original_vs_retrieved}
\end{figure}

\begin{figure}[htbp]
    \centering
    \begin{subfigure}{0.23\textwidth}
        \includegraphics[width=\textwidth]{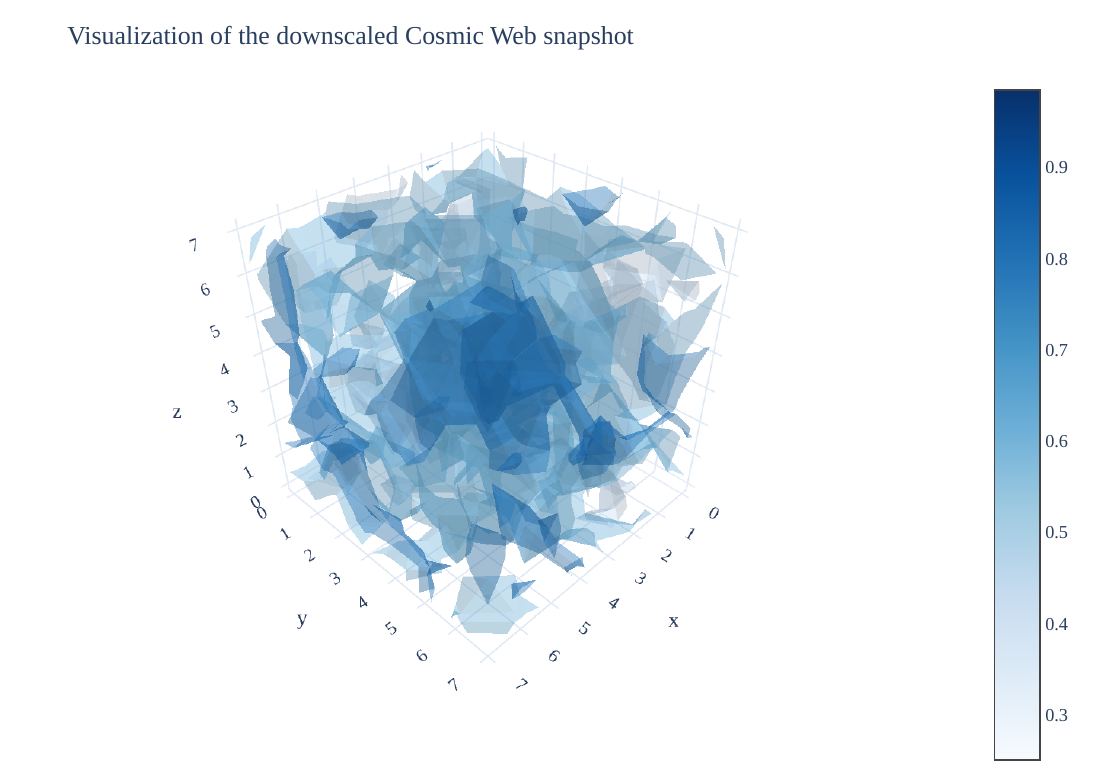}
        \caption{original}
        \label{fig:mfrqi_original_snapshot_e_normalized_median}
    \end{subfigure}
    \begin{subfigure}{0.23\textwidth}
        \includegraphics[width=\textwidth]{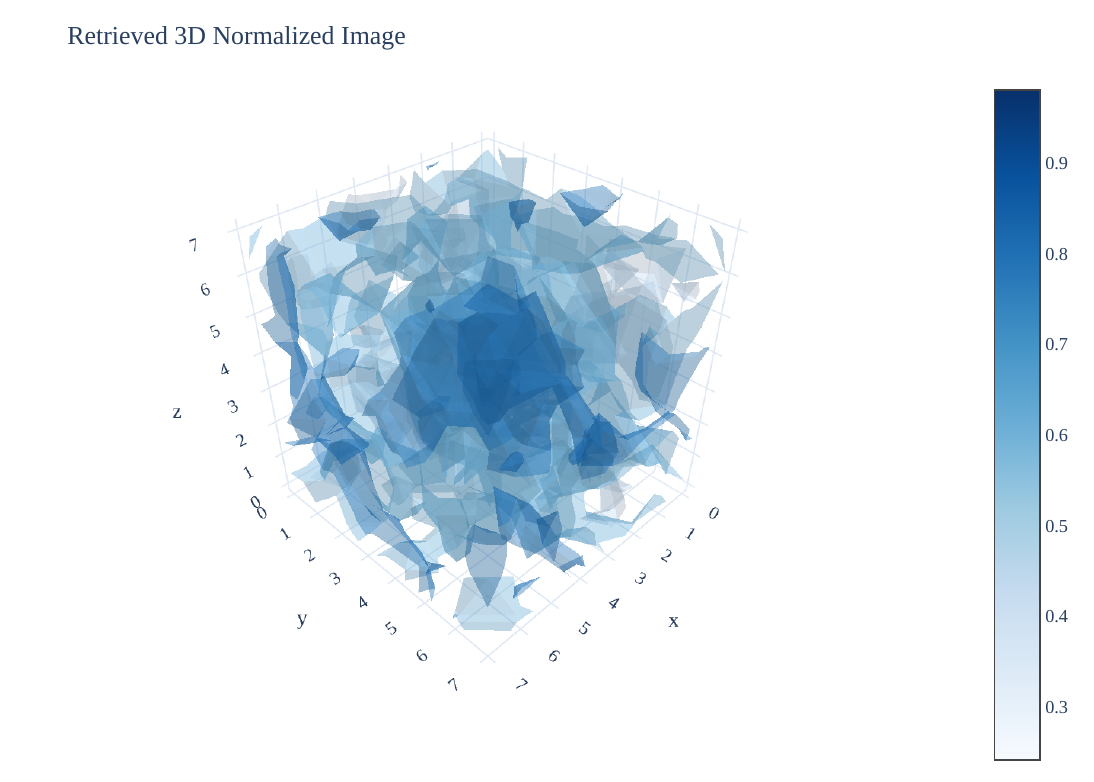}
        \caption{retrieved}
        \label{fig:mfrqi_retrieved_3d_normalized_image}
    \end{subfigure}

    \caption{Comparison of snapshots normalized with \cref{eq:normalizaion_e_median}: (a) original vs (b) retrieved snapshot generated using GEQIE framework (PCC=0.999).}
    \label{fig:mfrqi_original_vs_retrieved_normalized}
\end{figure}

Comparison of the original down-scaled snapshot with the retrieved snapshot is shown in \cref{fig:mfrqi_original_vs_retrieved}. The retrieved images show a strong resemblance to the original, with the calculated PCC value of 0.995. The Comparison of intermediate stage -- normalized arrays, that are encoded with GEQIE and later directly retrieved, is shown in \cref{fig:mfrqi_original_vs_retrieved_normalized}, and the intermediate PCC value is 0.999.

\section{Discussion}\label{sec:discussion}

As mentioned in \cref{sec:introduction}, two frameworks were introduced to unify various quantum image representation methods, i.e., NGQR and QPIXL. Their closer examination reveals that each framework imposes specific constraints on how quantum image representations can be constructed and processed. NGQR was proposed as a gate-oriented framework providing a formal structure for quantum image manipulation; however, its authors have not reported a publicly available source-code implementation. In contrast, QPIXL is delivered as a fully implemented framework based on the QCLAB++ library and is primarily focused on circuit-level optimization aimed at reducing quantum circuit complexity rather than unifying encoding methods.     

In this work, the proposed GEQIE framework approaches quantum image representation from a higher level of abstraction, beyond the explicit specification of individual quantum gates. By operating at the representation level and relying on a transpilation stage, GEQIE enables hardware-agnostic adoption, independent of specific quantum device constraints. Moreover, GEQIE is implemented in Python and built on the Qiskit ecosystem, which, unlike QCLAB++, provides direct access to IBM's real quantum hardware. Those facts significantly enhance the practical applicability of the framework for both experimental validation and near-term quantum computing platforms.

The benchmarking results provide a clear indication of the framework's operability and show how easily GEQIE can be integrated into more sophisticated experiments. Results show that some methods (IFRQI, NEQR, MCQI, QRCI, QUALPI) that use more qubits for color encoding are more resilient to noise, even under extreme conditions. However, this comes at the cost of high computational resources for circuit simulation. Other methods, such as FRQCI, suffer from high retrieval error even in the ideal simulation regime. It should be possible to reduce this error using another color encoding method, for example, by arranging the channel bits sequentially, starting with the most significant bit -- this way, at least the rough higher-order information is preserved, shifting the sampling error onto the least significant bits. This approach can be examined in future work, especially regarding the Cosmic Web encoding, which will require more channels for better information retention. Nevertheless, there exist better methods, such as MCQI, NCQI, or QRCI, that deliver good results out of the box under the assumption of fault-tolerant computations.

The toy example of Cosmic Web encoding shows the applicability of the framework in other fields of research. Despite being only a showcase, we have identified its potential, and this area may benefit from deeper research involving dark matter evolution simulations. The simulations require more data to be stored, for example, the object's velocity, to track the object's inertia over simulation frames correctly. One solution is the use of multichannel encoding, such as MFRQI, modified by adding more channels, or QRCI or NCQI, extended to three dimensions.

The important issue we have encountered is the cost of simulating the circuit. Some authors, like \cite{isakov2021simulations}, suggest simulating quantum circuits on the Google Cloud Platform using larger virtual machines or GPUs, which will allow up to 40 qubits. However, the underlying simulation framework differs from the one used in the current implementation of GEQIE. It is worth considering extending the simulation backend to Cirq \cite{cirqdevelopers2025cirq} and qsim \cite{qsim2025qsim}.

\section{Conclusions}\label{sec:conclusions}

The presented framework can encode quantum images from the Quantum Lattice family, enabling easy addition of new methods and quick execution of experiments involving existing ones. The current state allows any of the implemented quantum methods to be used out of the box and incorporated into more sophisticated experimental setups. As the software is available open-source, we want to encourage contributions of new methods.

The example of Cosmic Web image encoding shows that snapshot data can be encoded into a quantum state and retrieved with high quality. This experiment appeals as a promising source for more detailed and specialized research in this field.

The abstraction layer introduced over Quantum Image Encoding methods addresses the gap between researchers and current low-level quantum computing algorithms. It provides a new paradigm, opening the field as a high-level experimental framework rather than a hardware-centric discipline. By separating data representation from circuit-level implementation, the proposed approach enables domain researchers to engage with quantum methodologies without requiring detailed knowledge of gate-based design. In this way, the framework acts as a bridge between foundational quantum computing and applied research, supporting interdisciplinary experimentation and accelerating the transition of quantum computing toward practical, domain-oriented applications.

\section*{Acknowledgments}

The authors acknowledge that this paper is based on the results achieved within the OptiQ project. This project has received funding from the European Union’s Horizon Europe program under grant agreement No 101080374-OptiQ.

Additionally, the project is co-financed by the Polish Ministry of Science and Higher Education within the International Co-financed Projects program. The work is co-financed by the Silesian University of Technology (Poland) under grant number 02/080/BKM25/0057.

Disclaimer. Funded by the European Union. However, views and opinions expressed are those of the author(s) alone and do not necessarily reflect those of the European Union or the European Research Executive Agency (REA – granting authority). Neither the European Union nor the granting authority can be held responsible for them.

\bibliographystyle{quantum}
\bibliography{refs}

\end{document}